\documentclass[a4paper,preprint,aps]{revtex4}

\usepackage{graphicx}
\usepackage{dcolumn}
\usepackage{bm}

\newcommand{\eqa}{\begin{equation}}
\newcommand{\eqz}{\end{equation}}
\newcommand{\eqma}{\begin{eqnarray}}
\newcommand{\eqmz}{\end{eqnarray}}

\begin{document}
\title{The Role of the Basis Set: Assessing Density Functional Theory}
\author{A. Daniel Boese}
\affiliation{Department of Organic Chemistry,
Weizmann Institute of Science, IL-76100 Re\d{h}ovot, Israel}
\author{Jan M. L. Martin}
\affiliation{Department of Organic Chemistry,
Weizmann Institute of Science, IL-76100 Re\d{h}ovot, Israel}
\author{Nicholas C. Handy}
\affiliation{University of Cambridge, Lensfield Road, Cambridge CB2 1EW, UK}
\date{Draft version \today}
\smallskip
\begin{abstract}
\indent
When developing and assessing density functional theory methods, a finite basis set is usually employed.
In most cases, however, the issue of basis set dependency is neglected. Here, we assess
several basis sets and functionals. In addition, the dependency of the semiempirical fits
to a given basis set for a generalised gradient approximation and a hybrid functional is investigated.
The resulting functionals are then tested for other basis sets, evaluating their errors and
transferability.
\end{abstract}
\maketitle  
\section{Introduction}
In the past years, Density Functional Theory (DFT) has become a very important approach for
computational quantum chemistry. The Kohn-Sham implementation of DFT critically depends
on the quality of the exchange-correlation functional for its success. Recently,
various second-generation functionals (such as PBE\cite{PBE},mPW91\cite{mPW91},VSXC\cite{VSXC},
PBE0\cite{PBE0},PKZB\cite{PKZB},HCTH/93 and B97-1\cite{HCTH93},HCTH/120\cite{HCTH120},
HCTH/407\cite{HCTH407},
OPTX\cite{OPTX},B972\cite{B972},$\tau$-HCTH and its hybrid\cite{tHCTH},mPW1K\cite{mPW1K},B97\cite{B97},
and B98\cite{B98}) have been developed. These add to the numerous functionals that are
already available and commonly used in standard program packages (such as B88X\cite{B88X},
B3P91 and B3LYP\cite{B3P91},VWN\cite{VWN},P86\cite{P86},LYP\cite{LYP},P91X\cite{P91X}, and
P91c\cite{P91c}).
All of the functionals were developed from a wide variety of considerations, with most of them
focused on the exchange-correlation hole and employing different philosophies in their approximations.

Usually, after deciding upon the mathematical form of the functional, its parameters
have to be obtained. The latter is a difficult process where different
routes can be followed. Some of these functionals only use parameters that were
determined by considering known boundary conditions that the functional or density should
obey, and for properties of certain idealised systems like the uniform electron gas
\cite{PBE,mPW91,VWN,P86,LYP,P91X,P91c}; hence their parameters are mainly determined
by physical properties. The majority of commonly used functionals have been determined by fitting
at least one of their parameters to atomic or molecular data. A number of the latter functionals have been
generated using numerical DFT (i.e. basis set-free) methods 
\cite{B97,B98,B88X,B3P91} in their fits (although they in principle introduce numerical noise), but
most of the functionals have been fit to molecular sets with limited basis sets, usually of
triple-zeta\cite{VSXC,PBE0,PKZB,HCTH93,HCTH120,HCTH407,OPTX,B972,tHCTH} or even
double-zeta\cite{mPW1K} quality.

Many DFT users are thus overwhelmed by the sheer number of functionals and possibilities that
can be used, whereas with {\it ab initio} methods, the choices are clearly defined and mainly determined by 
a trade-off between rigor and computational cost. Very often because of sheer user inertia,
first-generation functionals are applied rather than the more accurate second-generation functionals.
Generally, a basis set of double- or triple-zeta quality is applied. Meanwhile,
systematic studies on the dependency of the basis set and functionals remain sparse
\cite{basquad,basquad2,basdip,basat1,basblyp,basfreq,basrev}. In addition, it is not a priori
clear that basis sets
optimised for wavefunction {\it ab initio} methods are the most optimal choice of DFT. Recently,
new basis sets especially optimised for DFT \cite{Jensen1,Jensen2,GSAW,DFO} have been proposed.
However, we have to keep in mind that in these cases the basis sets have been optimised for specific
functionals, like BLYP\cite{Jensen1,Jensen2} or the Local Spin Density  Approximation (LSDA)
\cite{GSAW,DFO}, further adding to the confusion. In sum, we have a plethora of functionals
developed for specific basis sets and additionally, a variety of basis sets developed for specific
functionals. In the former case, the question arises which basis sets can and should be used for
developing new functionals, and which basis set should then be employed when applying them. This is
a major consideration in functional development, since the question remains: with which basis set the
error of the basis set becomes comparable to that of the functional? Even if part of the basis
set error is absorbed into the parameters of the functional, the question remains of how transferable
such a functional will be to other basis sets- be they further from, or nearer to the infinite
basis set limit.

In this contribution, we will evaluate and compare various functionals using basis sets of
double-zeta {\it sp} to quadruple-zeta {\it spdfg} quality. In addition, we will fit a functional,
in this case the Hamprecht-Cohen-Tozer-Handy (HCTH)
functional \cite{HCTH93}, to slightly modified G2-1\cite{G2-1} and G3 sets of molecules\cite{G3}.
These sets are denoted as the 147 \cite{HCTH120} and 407 Sets \cite{www}.
All these fits will be carried out for several basis sets of double-zeta, triple- and quadruple-zeta
quality. 
The same fitting procedure was employed for the functional \cite{HCTH93,HCTH120,HCTH407,
HCTH407+} using the TZ2P basis set.

HCTH is a post-local spin density approximation (post- LSDA) functional, meaning that it factorises
the LSDA functional forms ($F_{LSDA}$), which can be found elsewhere \cite{PW91}:
\begin{eqnarray}
E_{xc} = \sum_{\gamma=x,c_{\sigma\sigma},c_{\alpha\beta}}\!\!\!\!\!\!\!E_{\gamma} = \sum_ {\gamma}
\sum_{q=0}^{m} c_{q,\gamma}
\int \!\!F_{LSDA,\gamma}(\rho_{\alpha},\rho_{\beta})
f_{\gamma,q}(\rho_{\alpha},\rho_{\beta},x_{\alpha}^2,x_{\beta}^2) d{\bf r}
\end{eqnarray}
where, $f_{\gamma,q}$ denotes the perturbation from the uniform electron gas if $c_{0,\gamma}=1$.
\begin{eqnarray}
f_{\gamma,q}&=&u_{\gamma}^q=\left(\frac{\theta_{\gamma\sigma}x_{\sigma}^2}{1+\theta_{X\sigma}x_{\sigma}^2}\right)^q
\end{eqnarray}
$x_{\gamma}$ is closely related to the reduced density gradient, and $\theta$ are fixed coefficients, which
have been fit to atomic data \cite{B97}.
\begin{eqnarray}
x_{\sigma}^2=\frac{(\nabla\rho_{\sigma})^2}{\rho_{\sigma}^{8/3}}
\end{eqnarray}

When employing the form in Eqn. 1 up to fourth order in $m$, we obtain 15 linear coefficients
(because of exchange, like-spin and opposite-spin correlation), which
are easily parameterised by minimising $\Omega$:
\begin{eqnarray}
\Omega&=&\sum_{m}^{n_E} w_{m} (E_{m}^{exact}-E_{m}^{K-S})^2 + \sum_{l,X}^{n_G} w_{l,G} 
\left(\frac{\partial E_{l}^{K-S}}{\partial X}\right)^2\nonumber\\
&&+ \sum_{j,\sigma}^{n_v} w_{j,v} \int (v_{j,\sigma}^{ZMP} + k_{j,\sigma} - v_{j,\sigma}^{K-S})^2 \rho_{j,\sigma}^{2/3}\!\! d{\bf r}
\end{eqnarray} 
The three summations correspond to errors of the energies, gradients and exchange-correlation potentials of
each molecule, respectively. In all cases, K-S denotes the calculated property; hence we have the
energy difference between
the exact and calculated energy in the first sum. In the second sum the exact gradients (at
equilibrium geometry) should be zero. In the final term, we fit to the 
exchange-correlation potentials determined by the Zhao-Morrison-Parr method \cite{ZMP}
from high-level {\it ab initio} densities, which are shifted 
by a constant $k$ because of the effects of the quantum-mechanical integer discontinuity.
All these contributions need to be weighted by appropriate weights $w$, which have
been determined and reported in previous papers \cite{HCTH407}. The weights $w$ consist of several
factorised weights making contributions for each molecule in order to ensure a balanced functional.

In the next section, we will refit the HCTH generalised gradient approximation (GGA) functional
to numerous basis sets. In addition, we will assess the performance of these functionals with
basis sets other than those used for the parameterisation. In the third section we will apply
the same procedure to hybrid
functionals, focusing on the amount of exact exchange needed for those functionals depending on
the basis set. 
In the last section, we will assess several functionals with two triple-zeta basis sets (namely,
TZ2P\cite{TZ2P} and cc-pVTZ\cite{PVXZfirst,PVXZsecond}).

\section{GGA Functionals and Basis Sets}

For all calculations, we used the Cadpac suite of programs \cite{Cadpac}, using a standard 'high'
grid for the density functional calculations. In the basis set evaluation, we used the
3-21G\cite{321G1,321G2}, 6-31G\cite{631G1,631G2}, 6-311G\cite{6311G1,6311G2} (the latter two with various
combinations of diffuse and polarisation functions), DZP, TZ2P\cite{TZ2P}, DFO\cite{DFO}, cc-pVDZ,
aug-cc-pVDZ, cc-pVTZ, aug-cc-pVTZ and cc-pVQZ basis sets\cite{PVXZfirst,PVXZsecond}. Some of the 
basis sets specifically constructed for density functionals discussed in the introduction are not
available (yet) for second-row atoms \cite{Jensen1,Jensen2,GSAW} and thus could not be used.

In Table \ref{tab1}, we compare the performance of the functionals with the basis set with which they
were fit.
HCTH/147@3-21G, for example, denotes an HCTH functional fitted to the 147 set with a 3-21G basis set.
The first column contains the RMS energy error
of the functional for the atomisation and dissociation energies, electron affinities, proton affinities,
and ionisation potentials in the 147 Set. The second column displays the sum of all the
gradients calculated at the equilibrium geometry in atomic units. The gradients have been shown to
correlate with the accuracy of the bond distances and angles obtained\cite{HCTH407,tHCTH}. The third
column is the square of all the errors in the exchange-correlation potential contributions. All
three errors are actually fitted according to equation 4. In the fourth column, we evaluate
$Q$, which is the sum of all three weighted errors with uniform weights rather than $\Omega$:
\begin{eqnarray}
Q&=&750\times\sum_{m}^{n_E}(E_{m}^{exact}-E_{m}^{K-S})^2 + 500\times\sum_{l,X}^{n_G} 
\left(\frac{\partial E_{l}^{K-S}}{\partial X}\right)^2\nonumber\\
&&+ \sum_{j,\sigma}^{n_v} \int (v_{j,\sigma}^{ZMP} + k_{j,\sigma} - v_{j,\sigma}^{K-S})^2\rho_{j,\sigma}^{2/3}\!\! d{\bf r}
\end{eqnarray} 
Since all three contributions are important and the GGA functionals have been fit to  a value similar to
$Q$, it is probably
the most important diagnostic. Nevertheless, as only the first two sums are directly apparent from
the energy calculations, we also evaluate Q$_1$, which excludes the potential:
\begin{eqnarray}
Q_1&=&750\times\sum_{m}^{n_E}(E_{m}^{exact}-E_{m}^{K-S})^2 + 500\times\sum_{l,X}^{n_G} 
\left(\frac{\partial E_{l}^{K-S}}{\partial X}\right)^2
\end{eqnarray} 
We have to bear in mind that both the ``exact'' exchange-correlation potentials and densities were
calculated using the TZ2P basis set. This, however, does not necessarily imply that the HCTH/147@TZ2P
functional has a distinct advantage when we fit to these quantities since basis set
convergence with angular momentum is much slower in {\it ab initio} methods than for density functionals.
Nevertheless, the density calculations using the Brueckner Doubles method include core correlation.
This might suggest that the TZ2P basis set has a slight advantage over the other basis sets
when fitting it to ZMP potentials that have been obtained from BD densities\cite{BD}. Unlike the
cc-pVTZ and 6-311G basis sets, the TZ2P basis set is of triple-zeta quality in the inner-shell orbitals.

Ignoring for the moment the Pople basis sets that just differ in the polarisation component on the
hydrogen, the following ordering of HCTH/147@ in $Q$ is observed:

TZ2P $<$ 6-311+G(3df,2pd) $\approx$ 6-311+G(3d,2p) $\approx$
aug-cc-pVTZ $\approx$ cc-pVQZ $\approx$ 6-311+G(2d,2p) $<$ cc-pVTZ $<$ 6-311G(2d,p) $\ll$ DFO2

All basis sets are ordered by their errors (unlike Table \ref{tab1}), with $<$ and $\ll$ corresponding
to a difference of more than 5\% or 20\% between the functionals, respectively. From Table \ref{tab1},
the HCTH/147@TZ2P functional clearly yields the best overall results. The Pople large basis sets also
exhibit surprisingly low errors, although some polarisation and diffuse functions
are needed to achieve this performance. Interestingly, Dunning's correlation consistent basis sets,
which were optimised at the CISD level, show high errors despite having more basis functions.
Even at the quadruple-zeta level, this ``basis set functional'' yields higher errors than both
HCTH/147@TZ2P and HCTH/147@6-311+G(3d,2p). HCTH/147@DFO2, whose basis set was developed specifically
for density functionals, shows an extraordinarily poor performance, probably due to built-in constraint
of DFO2 that the different angular momentum functions share the same exponents.

When excluding the potentials at the triple-zeta level, $Q$ changes to $Q_1$, and the ordering of
HCTH/147@basis set becomes:

6-311+G(3df,2pd) $\approx$ 6-311+G(3d,2p) $<$ cc-pVQZ $\approx$ TZ2P
$\approx$ 6-311+G(2d,2p) $<$ aug-cc-pVTZ $\approx$ cc-pVTZ $<$ 6-311G(2d,p) $\ll$ DFO2

Here, the HCTH/147@TZ2P functional exhibits a somewhat higher error compared to the HCTH/147@6-311+G(3d,2p)
functionals for Q$_1$, yet still has an error comparable to the only quadruple-zeta basis set
tested, HCTH/147@cc-pVQZ. The latter basis sets yield only slightly lower
errors than when fitting to a TZ2P basis. For the double-zeta quality basis sets, the energies and
gradients have much larger error contributions to $Q$ than the potentials, hence the exclusion of
the latter has no impact on the ordering in the HCTH/147@ functionals:

6-31+G(2d,p) $<$ 6-31+G** $<$ DFO1 $\approx$ 6-31G(2d,p) $<$
aug-cc-pVDZ $<$ 6-31G** $<$ 6-31G* $<$ DZP $\ll$ cc-pVDZ $<<<$ 6-31G $<<<$ 3-21G
 
As for the double-zeta basis sets, the unpolarised basis sets are clearly not very
useful for the purpose. HCTH/147@DFO1 and the functionals fitted to the Pople basis sets yield
the lowest errors. The former should not be surprising as it was developed specifically for DFT
whereas the Pople basis sets were developed for Hartree-Fock. Nevertheless, with the
exception of the HCTH/147@6-31+G(2d,p) functional, the double-zeta ``basis set functionals'' yield
much higher errors than those obtained by the triple-zeta quality basis sets. {\it Hence,
if computationally feasible, basis sets of triple-zeta quality are preferable over basis sets of
double-zeta quality when doing calculations employing DFT.} Overall, and consistent with earlier
observations made by Jensen \cite{Jensen2}, Dunning's basis sets do not seem to be an optimal
choice for density functional calculations. On the other hand, diffuse functions are more important
with the Pople basis sets, in that they provide a significant error reduction. A detailed analysis of
the results shows that predominantly the total atom energies, atomisation energies and electron
affinities of anions are affected, with the error for the latter increasing by more than a factor
of two when omitting diffuse functions. In comparison, with the HCTH/147@cc-pVTZ functional the error
of the atomisation energies and ionisation potentials of the anions decreases by less than 20\% when
diffuse functions are included in the basis set. This is a general indication that despite their
limited usefulness for DFT, Dunning's basis sets are more ``balanced'' than the 6-31 family of basis
sets with respect to diffuse functions. The HCTH/147@TZ2P functional yields a similar error as the
HCTH/147@cc-pVTZ basis set for the anions, suggesting that diffuse functions might reduce its error
further. Changing the basis functions on the hydrogen atom, as has been suggested in a
prior assessment of different density functionals \cite{CH,CHcomment}, worsened the performance
of the HCTH/147@TZ2P ``basis set functional'' in all tests.

Having obtained a number of functionals, comparing their coefficients in Table \ref{tab2} 
yields further insight. Here, only the HCTH/147@3-21G functional obtained from
fitting to the split-valence 3-21G basis set differs blatantly from the other functionals.
Both the HCTH/147@DZP and HCTH/147@aug-cc-pVDZ functionals have large zeroth-order like-spin
correlation coefficients ($c_{C\sigma\sigma,0}$) probably thus compensating for the basis set error.
Note that this coefficient also increases when going
to larger molecules \cite{HCTH407}. The HCTH/147@6-31+G** coefficients are closer to the ones obtained
by the triple-zeta basis sets than the other double-zeta quality ``basis set functionals'', confirming its
lower basis set error.
No clear conclusions on this effect can be made with the triple- and quadruple-zeta functionals,
since both the HCTH/147@TZ2P (including the potential) and HCTH/147@6-311+G(3d,2p) functionals yield
the lowest errors and not the HCTH/147@cc-pVQZ functional. The latter basis set has supposedly the
lowest basis set error compared to the basis set limit \cite{Jensen2}. The exchange coefficients
($c_{X\sigma,n}$ with m going from 0 to 4) of all the functionals at this level of basis set
quality seem to be similar for the different functionals reported in Table \ref{tab2}.
In contrast, the correlation parameters exhibit a wider variation upon switching between the
fitting sets with 93, 147 or 407 systems \cite{HCTH407}. The effect
of the diffuse functions on the basis set going from the HCTH/147@cc-pVTZ
to the HCTH/147@aug-cc-pVTZ functional is not as significant as the change between the different
``basis set functionals'' of triple-zeta quality. Nevertheless, the coefficients still differ by a
significant amount, suggesting that convergence towards the basis set
limit has not been achieved. However, if similar basis set errors are absorbed into the
parameterisation of functionals fitted to different triple and quadruple-zeta basis sets, the
functionals are expected to be fairly transferable. In addition, the functional error
is only very slightly influenced by changing some of the functionals coefficients. We have to
note that using functionals with quadruple-zeta quality or higher is not useful, since
the basis set error is probably about a magnitude lower than the functional error itself.

In Tables \ref{tab3} to \ref{tab6} we investigate the transferability of the ``basis set
functionals'' in more detail. Here, all the new functionals developed are evaluated with the
6-31+G**, aug-cc-pVDZ, 6-311+G(3d,2p) and TZ2P basis sets.
The results for the 6-31+G** basis for selected functionals that are listed in
Table \ref{tab1} are shown in Table \ref{tab3}. As expected, the HCTH/147@3-21G and HCTH/147@6-31G 
functionals are clearly insufficient to describe the properties investigated.
Reasonably low errors are obtained with all functionals being fit to the 6-31G* basis
set or larger. Interestingly, in the case of the 6-31+G** basis set, only the HCTH/147@DZP
and HCTH/147@aug-cc-pVDZ functionals show $Q_1$ errors below 40 a.u., affected
mainly by the gradient error contribution. In comparison (Table \ref{tab1}), the functional optimised
for this basis set has a Q$_1$ value of 32.5 a.u.. For all the triple-zeta quality functionals,
the RMS energy error is lower, but the gradients (geometries) counterbalance this effect. All of
these ``basis set functionals'', with the exception of HCTH/147@6-311+G(3df,3pd) and
HCTH/147@aug-cc-pVTZ, give very similar $Q_1$ errors, with RMS energy errors of 
6.7 $\pm$ 0.3 kcal/mol and a gradient error around 3.6 a.u.. Overall, the triple-zeta and
quadruple-zeta quality functionals vary by at most 20\% in both errors.

The evaluation of the functionals with the aug-cc-pVDZ basis set yields similar results (Table \ref{tab4}).
Of course, the overall error $Q_1$ is higher, as is evident from the larger error obtained by the
HCTH/147@aug-cc-pVDZ functional compared to the HCTH/147@6-31+G** functional displayed in
Table \ref{tab1}. The gradient errors are especially affected. Only the double-zeta quality
``basis set functionals'' yield errors that can be compared to the $Q_1$ value of 42.4 a.u.
obtained with the HCTH/147@aug-cc-pVDZ functional (Table \ref{tab1}).
Here, the HCTH/147@6-311+G(3d,2p), HCTH/147@6-311+G(3df,2pd), HCTH/147@TZ2P, HCTH/147@aug-cc-pVTZ
and HCTH/147@cc-pVQZ functionals again yield higher errors, mainly because of the gradient
(and partly due to the energy) contributions to the error $Q_1$. Thus, the functionals fit to
the double-zeta quality basis sets give the lowest errors when applied to {\it other} basis sets
of double-zeta quality.

If we evaluate the functionals with triple-zeta quality basis sets, we observe, as expected, that their
errors are a lot closer to the lowest error possible obtained by the basis sets to which the functionals
were fit.
When using the 6-311+G(3d,2p) basis set, the double-zeta ``basis set functionals'' yield the lowest
gradient errors, although again this is only achieved when including diffuse diffuse functions.
As for the other entire basis set evaluations in Tables \ref{tab3} to \ref{tab6}, the double-zeta
basis sets give a much lower error for the gradients, but not for the energy. This is an interesting
phenomenon; the same observation was made in a different context\cite{Menconi} when GGA functionals were
developed solely for the description of accurate structures and frequencies\cite{Menconif}. Hence, it is
probably a lot harder to develop a single functional for both accurate geometries and energies rather than one for
each individual property. However, the former approach of separating the calculations can lead to
other problems, namely that energetic properties will be calculated at non-equilibrium structures.
The HCTH/147@cc-pVQZ, HCTH/147@cc-pVTZ and HCTH/147@aug-cc-pVTZ functionals again yield higher errors than
the other functionals developed with a triple-zeta quality basis set. Noteworthy is the low error of the
HCTH/147@TZ2P functional when applied to the 6-311+G(3d,2p) basis set, resulting in a $Q_1$ value of 15.5
compared to the minimum value of 15.0 a.u. obtained by the HCTH/147@6-311+G(3d,2p) functional. This is even
lower than the error of the HCTH/147@6-311+G(3df,2pd) functional obtained with the 6-311+G(3d,2p) basis set.
This emphasises the transferability of the functionals that have been fitted to the higher
basis sets, since the variance is less than 10\% on $Q_1$ and on the RMS energy error and the gradient errors
(with the exception of the HCTH/147@cc-pVTZ and HCTH/147@6-311+G(2d,p) functionals for the RMS energy error).
A number of the functionals parameterised for the triple-zeta basis sets yield a lower error when
evaluating $Q_1$ with the 6-311+G(3d,2p) basis set rather than for the basis set to which they were fit
(compare to Table \ref{tab1}). This indicates that a large amount of the remaining basis set errors
absorbed into these functionals is of a similar magnitude.

In Table \ref{tab6}, the performance of the different ``basis set functionals'' with the TZ2P
basis set is shown. All errors are slightly larger (by about 10\%), but the trends are again the
same. The HCTH/147@DZP basis set
functional now gives a lower error than the HCTH/147@6-31+G** functional, but only by a small
margin. Again, the functionals  fitted to Dunning's basis sets of double-zeta quality yield larger
errors than the other functionals fitted to basis sets of double-zeta quality. Among the triple-zeta
level ``basis set functionals'', the errors vary by only 10\%, supporting the observations made for the
6-311+G(3d,2p) basis set. When evaluating the HCTH/147@3-21G and HCTH/147@6-31G functionals in
Tables \ref{tab4} to \ref{tab6}, the former functional has a lower gradient error.
Hence, polarisation functions are essential when calculating geometries, reducing the gradient
errors by a factor of two and more.

In summary, the triple-zeta ``basis set functionals'' are transferable between each other,
indicating that when fitting using basis sets of this quality, the basis set error absorbed in the
parameterisation does not play a role.
The variance of 10\% or less when evaluating one functional fit with a certain basis set with a different
one is probably not important. This has further implications in the use of density functionals. {\it Since
the errors do not change significantly when going to higher basis sets, the triple-zeta basis set
level is likely to be sufficient for use in density functional calculations.} With still larger basis
sets, the basis set truncation error will ``drown in the noise'' that is the inherent error of the
functional itself.
In general, Dunning's correlation-consistent basis sets developed for correlated {\it ab initio} methods
yield higher errors than the various Pople basis sets or TZ2P. When we
investigated this behaviour in more detail, we found that the inclusion of core-valence basis functions
\cite{CWNZ,MartinCV} only lowers the RMS energy error by 0.1 kcal/mol. Inclusion of an additional tight
d-function\cite{WilsonD}, which has been shown to be important for second-row
elements\cite{SO2martin}, reduces this error by a further 0.3 kcal/mol. Furthermore, these contributions
lower the sum of the gradient error by another 0.1 a.u. If we assume that the same contributions were added
to the HCTH/147@aug-cc-pVTZ functionals results, we would probably arrive at a $Q$ value close to
the lowest value obtained.
Thus, although several enhancements for correlation-consistent basis sets could be introduced,
extensive complements of higher-angular momentum functions, in particular, do not appear to be necessary.
Out of all basis sets tested, the HCTH functional fit to the 6-311+G(3df,2pd) basis set gives the lowest
energies and gradients errors, and the HCTH functional fit to the TZ2P basis set yields the lowest
errors for the quantity $Q$, which includes energies, gradients and potential points. We believe that this
is due to correlation effects described by the exchange-correlation potential in the core that
cannot be adequately described by the 6-311+G basis sets.

\section{Hybrid Functionals and Basis Sets}

The method applied in the last section to the GGA functionals is now used for hybrid functionals
to assess the different basis sets. There is, however, one complication: the varying amount of exact
Hartree-Fock exchange. Here, we fitted hybrid functionals to the 6-31+G**, aug-cc-pVDZ,
6-311+G(3d,2p), TZ2P and aug-cc-pVTZ basis sets, using the abovementioned procedure. While the
inclusion of the exchange-correlation potential points into the fits for hybrid functionals is
possible\cite{B972}, it is not clear if the overall performance of these functionals
is generally better\cite{anharm}, as we will also see in the final section. Therefore, we restrict
ourselves to fitting to gradients and energies only (similar to the determination of $Q_1$ in
the last section, see Eqn. 6). All functionals were again fit to the 147 systems, but
with m=2 in the power series in Eqn. 2. This cut-off in the power series expansion yields the B97-1
form rather than the HCTH form. Here, 9 linear coefficients are fit instead of the
15 in the GGA. The amount of exact exchange was varied over a range from 0\% to 50\%
in order to determine the minimum.

Before discussing the outcome of the results of the hybrid functionals, it is worth comparing
the values at 0\% Hartree-Fock exchange to the functionals obtained in the previous
section. The difference between them is the fit to exchange-correlation potentials, which will
raise the $Q_1$ error for the GGA functionals in contrast to the ``hybrid'' functional
at 0\%. Of course, another discrepancy is the different number of coefficients.
The newly obtained GGA functionals are displayed in Table \ref{tab7}, and can be compared (with
the differences mentioned) to those in Table \ref{tab1}. For the TZ2P, 6-31+G** and
aug-cc-pVDZ basis sets, the higher-order coefficients seem to change very little (see table
\ref{tab1}). While the energy error increases going from m=2 to m=4, the gradient error
decreases, with $Q_1$ close to the errors when
fitting only to gradients and energies with m=2. This has been previously reported\cite{CH,B97} and
has led to the conclusion that it is unnecessary to include orders higher than 2 in the power series.
Here, mainly the error in the potential is affected as an additional calculation with the TZ2P basis
set and the variable $Q$ shows. In this case, $Q$ for the HCTH/147 with m=4 form yields 37 a.u., 
compared to 51 a.u. for the B97-1/147 form with m=2, hence the overall $Q$ value is raised
by 40\%. Comparing Tables \ref{tab1} and \ref{tab7}, we realise that {\it not only is the functional somwehat
dependent on the basis set for which it was parameterised, but also the basis set dependence itself is
dependent on the class of functional used}.

\begin{figure}
\vspace{3cm}
\includegraphics[width=10cm]{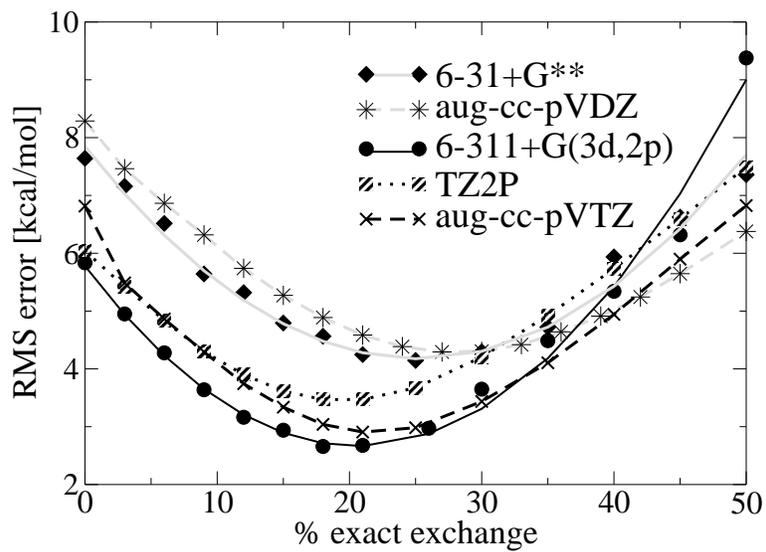}
\caption{\label{fig1}Boese et al, Journal of Chemical Physics}
\end{figure}

\begin{figure}
\vspace{3cm}
\includegraphics[width=10cm]{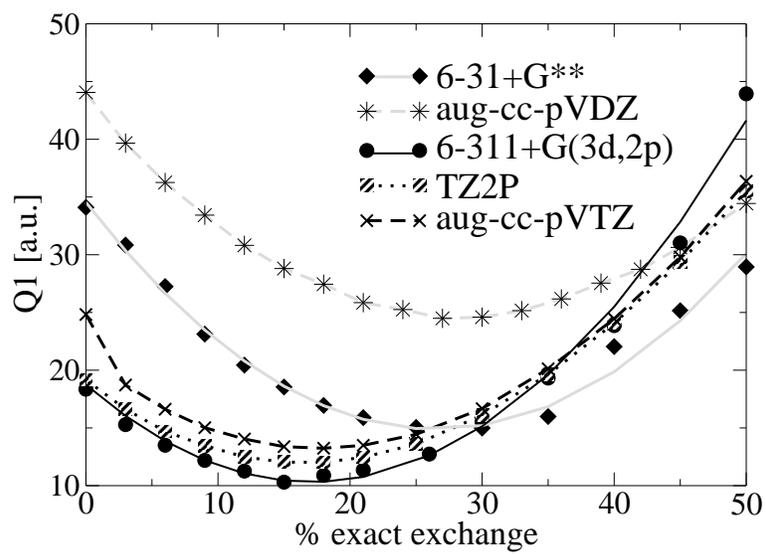}
\caption{\label{fig2}Boese et al, Journal of Chemical Physics}
\end{figure}

Let us now return to the hybrid functionals. Since the error of the resulting
functional also correlates slightly with the starting guess used, the points will not
necessarily fit a curve. Unlike the GGA functionals where we can determine the starting
guess by fitting to exchange-correlation points, with hybrid functionals the initial coefficients
are, at best, educated guesses. Thus, we sometimes had to fit a curve through the points, and in
Figure \ref{fig1}, the RMS energy error is plotted as a function of amount of exact exchange for
various functionals. This plot shows a disturbing property of hybrid functionals:
{\it the amount of exact exchange in the functional depends on the basis set for which it is fit.}
All hybrid density functionals known have their exchange coefficient fit to a specific basis
set (or are using numerical DFT), hence the variation of the exact exchange, in the range of
15\% to 25\% might well be due to basis set effects.
In the case of the tested functionals, the optimised exact exchange-fractions for all three
triple-zeta basis sets
(obtaining B97-1/147@6-311+G(3d,2p), B97-1/147@TZ2P and B97-1/147@aug-cc-pVTZ) hover around 18\%. However,
fitting to a double-zeta basis set (B97-1/147@6-31+G** and B97-1/147@aug-cc-pVDZ) yields minima
located around 28\%.
In Figure \ref{fig2}, the value of $Q_1$ is displayed, exhibiting generally the same behaviour as the
energy in Figure \ref{fig1}. Since this is the value used in the fit (with more sophisticated weights),
it is more informative, albeit more abstract. Here, the differences between the basis sets, which
were already exhibited with the GGA functionals, can be seen again. While the B97-1/147@6-31+G**
and the B97-1/147@aug-cc-pVDZ functionals show similar minima in the energies, with the
B97-1/147@6-31+G** curve shifted a bit towards lower values. Still, the gradient error when
fitting to the 6-31+G** basis is much lower than with the aug-cc-pVDZ basis set. Similar behaviour is
observed for the basis sets of triple-zeta quality.
Only when the gradients are included does the B97-1/147@TZ2P functional yield a lower error for
its minimum than B97-1/147@aug-cc-pVTZ, with the values obtained by the B97-1/147@6-311+G(3d,2p)
functional marginally lower still. Detailed results of the minima are given in Table
\ref{tab7} displaying the lowest energy points calculated. The minima obtained when fitting
a separate curve through the points are at 29 and 28\% for the B97-1/147@6-311+G** and
B97-1/147@aug-cc-pVDZ ``basis set functionals'', and the minima for the B97-1/147@6-311+G(3d,2p),
B97-1/147@TZ2P and B97-1/147@aug-cc-pVTZ ``basis set functionals'' are at 16, 17 and 17\%, respectively.

\section{Assessment of Density Functionals}

In this section we assess several density functionals with two basis sets: cc-pVTZ and
TZ2P. The former basis set has been used in the construction of VSXC\cite{VSXC}, and the
latter in the parameterisation of the HCTH-type functionals\cite{HCTH93,HCTH120,HCTH407,B972,tHCTH}.
For this evaluation, we use the large 407 Set\cite{www}, again comparing RMS energy, gradient
and $Q_1$ errors. A similar comparison for most of the functionals evaluated has been done before,
using the TZ2P basis set and the much smaller sets of 93 and 147 molecules. Thus, we can assess if
the results will remain transferable between the G2-1 and G3 sets\cite{CH}. Table \ref{tab8}
shows all results obtained.
\begin{itemize}
\item The simplest density functional method, LSDA in its VWN parameterisation, is already
a vast improvement over the Hartree-Fock method. The LSDA geometries are surprisingly
accurate, even more so when comparing the gradient error to a vast number of GGA and meta-GGA
functionals.
\item For molecular systems, PKZB yields no clear improvement over PBE, although it was
developed as an improvement over PBE including a semi-empirical fit and an extra variable,
the kinetic energy density $\tau$. Both functionals
give RMS energy errors close to 20 kcal/mol. In addition, we would expect geometry errors
similar to the ones obtained by the LSDA method. We would discourage the use of these functionals
for the calculation of both thermochemical data and geometries.
\item mPW1K is a hybrid functional with a large percentage of exact exchange, developed for
accurate reproduction of reaction barriers, reducing the error of B3LYP for this property by
about 50\%\cite{mPW1Keval}. However, its performance for minimum geometries and energetics of stable
molecules is the worst of all hybrid functionals tested and even standard GGA's yield lower errors.
\item PW91PW91, the original GGA proposed by Perdew, yields slightly lower errors than PBE,
that largely come from the evaluation of the gradients.
\item When replacing the PW91 exchange functional with mPW91, the RMS energy error is reduced, yet the
gradient error increases.
\item The BP86 GGA functional, albeit an improvement over these methods, still gives a
$Q_1$ error almost three times larger than the best functionals available. The RMS energy error is
lowered by 5 kcal/mol when using the cc-pVTZ basis set instead of TZ2P; the gradient error however
increases.
\item BLYP is one of the more accurate functionals, but generally overestimates bond distances, and
its gradient error is extremely large. This leads to RMS errors in bond lengths that are almost
twice as high as for hybrid functionals\cite{HCTH407}. Nevertheless, it is still one of the
commonly used functionals, very often employed when calculating hydrogen bonds\cite{TBH} or in
Car-Parrinello Molecular Dynamics simulations of liquid water\cite{BLYPCP}. Its
RMS and gradient errors both increase when using the cc-pVTZ basis set. In contrast to the
smaller sets, the BLYP functional now clearly outperforms BP86 for the 407 set, whereas
this was not the case for the 93 set\cite{CH}.
\item The BPW91 functional uses Becke's exchange functional in combination with PW91 correlation, and
yields lower errors than the PW91 and mPW91 exchange functionals for the properties tested. Here, the
functional gives a performance similar to BLYP, perhaps slightly better. Again, its gradient error is
higher for the cc-pVTZ basis set.
\item PBE0 does not yield better energy predictions than the GGA functionals
mPW91PW91, BPW91 and BLYP. Interestingly, its gradient error is lower than the one obtained for
the most commonly used functional B3LYP, thus we would expect a pretty accurate description of
geometries and higher-order properties.
\item OLYP\cite{OLYP} is a clear improvement over BLYP for atomisation energies and reactions, and even
more so for molecular structures\cite{hoe,baker}. Its overall errors are almost comparable to those of the hybrid
functionals, much better than BLYP and BPW91. Here, an improved exchange functional OPTX is used instead
of Becke's exchange functional. However, its performance in hydrogen bonds is not as good as BLYP.\cite{BHunpub}
\item B98 does exactly the opposite of PBE0: We would expect its geometries to be further away from the
equilibrium than B3LYP, but its energy error is slightly decreased. In the overall
$Q_1$ evaluation, the error is very similar to the one obtained by the older B3LYP.
\item The B3LYP functional is probably the most widely used hybrid functional, and every new
functional is compared to its accuracy. The RMS error for the 407 Set is close to
10 kcal/mol, which is considerably large compared to a ``chemical accuracy'' of 2 kcal/mol. The gradient error is
again slightly increased for the cc-pVTZ basis set compared to TZ2P.
\item B97-2 is a reparameterisation of B97-1 including the ZMP potential points into the
fit of the exchange-correlation functional. Hence, we can expect its energy error to be worse than
B97-1, since the fit to an extra quantity usually worsens the performance to the energies. However,
its advantages over B97-1 still have to be established since even its gradient error is larger.
\item The performance of VSXC for the 407 set is very similar to B3LYP. This contrasts with the
bad performance of VSXC for the smaller 93 Set where it returned an error barely lower than the
BLYP functional\cite{CH}. Fitted to the cc-pVTZ basis set, it is the only functional yielding
similar errors for both basis sets. 
\item All HCTH functionals give errors which differ in $Q_1$ by less than 3\% for the TZ2P basis set 
and less than 8\% for the cc-pVTZ basis set. These functionals, while they are pure GGA functionals, yield
errors that can be compared to hybrid functionals like B3LYP for both the TZ2P and cc-pVTZ basis
sets. Here, the additional value added to the functional by reparameterising it to 147 or 407
systems is not obvious, although the HCTH/407 functional outperforms the other parameterisations
in the RMS energy error. The $Q_1$ value of HCTH/407 is the worst for the HCTH functionals 
with the cc-pVTZ basis set. The justification for the reparameterisation, making HCTH/407 a better
functional than HCTH/93 or HCTH/147, will become only visible when considering hydrogen bonds or
inorganic molecules\cite{TBH,Menconi} where error cancellation plays an important role.
Generally, this error cancellation cannot be expected in DFT methods. In some post-Hartree-Fock
methods like MP2 it is inherent, and hence the reparameterisation of the functionals remains important
to recapture such effects.
\item B97-1 is probably the best choice when it comes to using density functional hybrid
calculations, since it is already well tested and its calculated structures are similar to
those obtained by B3LYP (or even slightly better), and it outperforms B3LYP
by about a third in $Q_1$ when predicting energetic properties.
\item $\tau$-HCTH is for both basis sets an improvement over the HCTH functional showing that
the inclusion of the kinetic energy density can lower the error further. Unfortunately,
it lacks the performance of the HCTH/407 functional for weak interactions\cite{tHCTH,HCTH407+}.
\item The $\tau$-HCTH hybrid, when additionally including exact exchange, clearly yields the lowest
$Q_1$ value of all methods tested. Its error for the TZ2P basis set is 50\% lower than the one
obtained by B3LYP in addition to yielding a lower gradient error and structures\cite{tHCTH}
than B3LYP.
\end{itemize}
Summarising the results in table \ref{tab8}, by ranking all GGA functionals based on their 
$Q_1$ value, we get the following order (taking into account that some
functionals have been fit to one of the basis sets):

HCTH/147 $\approx$ HCTH/93 $\approx$ HCTH/407 $<$ OLYP $<$ BPW91 $<$ BLYP $<$ mPW91PW91 $<$ BP86
$<$ PW91PW91 $<$ PBE

For the meta-GGA's and hybrid functionals, the ordering is:

$\tau$-HCTH hybrid $<$ B97-1 $<$ $\tau$-HCTH $<$ VSXC $\approx$ B97-2 $\approx$ B3LYP $\approx$
B98 $<$ PBE0 $\ll$ mPW1K $<$ PKZB

Although RMS errors can give a lot of insight, the maximum errors are also considered important.
In light of this, we have examined in Table \ref{tab9} the number of molecules in the 407 set
for each functional that have large RMS energy errors (over 15 kcal/mol). As we have discussed in the
basis set evaluation, the cc-pVTZ basis set generally exhibits more outliers (molecules with
atypically large errors) than the TZ2P basis set. The results are very similar to the ones obtained in
Table \ref{tab8}, and compared to B3LYP, the best available functional (the $\tau$-HCTH hybrid)
cuts the number of outliers in half, yielding a considerable improvement.
With regard to the accuracy of density functional theory with all the evaluations which are done with the
407 Set, the RMS error of the functionals ranges between 6 ($\tau$-HCTH hybrid) and 21 kcal/mol (PBE),
with the most commonly used hybrid functional B3LYP yielding an error around 10 kcal/mol. The mean error
of these functionals is between 3.8 kcal and 15.6 kcal mol, with B3LYP yielding 6.2 kcal/mol. This can be
compared to empirical correction methods like G3\cite{G3}, which yield errors around 1 kcal/mol
for the G3 set that, however, does not include some of the molecules with the largest errors in our
407 set. Extrapolation methods like W2 \cite{W2} give mean errors around 0.5 kcal/mol for a considerably
smaller set similar to our 147 Set. Full MP2 yielded an RMS error of about 20 kcal/mol for the
407 Set using the TZ2P basis set. This generally places the accuracy of DFT between raw perturbation
theory results and coupled-cluster theory when calculating ground state energies.

\section{Conclusions}

From the above, we can make a number of observations concerning the use of DFT functionals and the basis
sets used with them.
All results regarding the basis sets are obtained by fitting functionals to basis sets, and
then evaluating their errors, thus the results are independent of the functional parameterisation.
For most of the properties investigated, including energies and gradients, the Pople basis
sets can be recommended.
Dunning's basis sets, fit to CISD, give considerably higher errors despite having
a larger number of basis functions. Although it yields a higher error than the 6-311+G(3df,2pd) basis
set for energies and gradients, the TZ2P basis set still gives the lowest overall error for the GGA
functionals when including the ZMP exchange-correlation potentials into the fit.
The basis set error, which might still be significant at the triple-zeta level, shows that the
functionals obtained by fitting to one basis set are transferable to other basis sets. Hence, it is
probably not important to reach the basis set limit when developing new density functionals, since
the overall DFT error is considerably larger.
Basis sets developed for DFT methods might alleviate this problem, but the problem remains as to
which functional to use for their development.
The same analysis for hybrid density functionals shows that the amount of exact exchange obtained is
dependent on the basis set itself. Whereas basis sets of double-zeta quality yielded minima
around 28\%, the triple-zeta basis sets evaluated had their minima around 18\%. Concerning the
difference between the hybrid ``basis set functionals'', the same conclusions as for the GGA functionals
can be drawn. The same trends are also visible when evaluating several other published functionals to
a large test set. We can deduce that several hybrid functionals, such as B97-1 and the
$\tau$-HCTH hybrid functional render errors which are significantly lower than the ones obtained
by B3LYP. As a pure GGA functional, the HCTH functional types give errors comparable to B3LYP for
the investigated properties.

Nevertheless, the accuracy of modern density functional theory cannot be compared to that of
{\it ab initio} extrapolation methods. The best functional tested yields an RMS energy error as
large as 6.3 kcal/mol for a large set of molecules, which is still far away from the
desired ``chemical accuracy'' of 1-2 kcal/mol.

\section{Acknowledgements}
A. D. Boese is grateful for financial support by the EPSRC, the Gottlieb Daimler- und
Karl Benz- Stiftung and the Feinberg Graduate School. This research was supported
by the Lise Meitner Center for Computational Chemistry and the Helen and Martin A. Kimmel
Center for Molecular Design.

\indent

\newpage
\pagestyle{empty}
\clearpage

\newpage
\pagestyle{empty}
\clearpage

\newpage
\clearpage
\noindent
Fig. 1:\\
RMS error (in kcal/mol) of the 147 Set with different hybrid ``basis set functionals''.\\

Fig. 2:\\
$Q_1$ error (in atomic units, see eqn. 6) of the 147 Set with different hybrid ``basis set functionals''.

\newpage
\clearpage
\begin{table}
\caption{The HCTH functional errors when fitted to the respective basis sets with the 147 Set.
The final column is the value of Q$_1$ which excludes the potential in the sum of Q.\label{tab1}}
\begin{tabular}{|l|l|l|l|l|l|} \hline
Property                  & RMS energy & Gradient     & Potential    & $Q$    & Q$_1$  \\ \hline
Functional                & [kcal/mol] & $\sum$[a.u.] & $\sum$[a.u.] & [a.u.] & [a.u.] \\ \hline\hline
HCTH/147@3-21G            & 46.89      & 7.50         &  45.4        & 925.1  & 879.6 \\ \hline
HCTH/147@6-31G            & 18.53      & 6.07         &  43.8        & 295.6  & 251.8 \\ \hline
HCTH/147@6-31G*           & 10.60      & 2.99         &  24.2        & 77.6   & 53.4 \\ \hline
HCTH/147@6-31G**          & 10.29      & 2.71         &  24.9        & 75.2   & 50.3 \\ \hline
HCTH/147@6-31G(2d,p)      &  9.70      & 2.11         &  27.4        & 63.9   & 36.5 \\ \hline
HCTH/147@DZP              &  9.25      & 2.99         &  29.2        & 88.6   & 59.4 \\ \hline
HCTH/147@DFO1             &  7.44      & 2.85         &  26.5        & 63.8   & 37.3 \\ \hline
HCTH/147@cc-pVDZ          &  9.51      & 4.04         &  26.9        & 100.6  & 73.7 \\ \hline
HCTH/147@6-31+G**         &  6.31      & 2.66         &  28.8        & 61.3   & 32.5 \\ \hline
HCTH/147@6-31+G(2d,p)     &  4.82      & 2.47         &  27.4        & 51.3   & 23.9 \\ \hline
HCTH/147@aug-cc-pVDZ      &  6.91      & 3.40         &  30.0        & 72.4   & 42.4 \\ \hline
HCTH/147@6-311G(2d,p)     &  7.00      & 2.48         &  21.4        & 49.1   & 27.7 \\ \hline
HCTH/147@TZ2P             &  4.89      & 2.11         &  19.4        & 36.5   & 17.1 \\ \hline
HCTH/147@DFO2             &  7.37      & 2.83         &  32.9        & 74.5   & 41.6 \\ \hline
HCTH/147@cc-pVTZ          &  5.92      & 2.27         &  24.2        & 45.4   & 21.2 \\ \hline
HCTH/147@6-311+G(2d,p)    &  4.82      & 2.48         &  21.4        & 42.8   & 21.4 \\ \hline
HCTH/147@6-311+G(2d,2p)   &  4.93      & 2.11         &  26.3        & 43.7   & 17.4 \\ \hline
HCTH/147@6-311+G(3d,2p)   &  4.59      & 2.05         &  25.8        & 40.8   & 15.0 \\ \hline
HCTH/147@6-311+G(3df,2pd) &  4.68      & 1.95         &  26.5        & 40.7   & 14.2 \\ \hline
HCTH/147@aug-cc-pVTZ      &  5.08      & 2.42         &  20.2        & 41.0   & 20.8 \\ \hline
HCTH/147@cc-pVQZ          &  4.96      & 2.10         &  25.4        & 42.2   & 16.8 \\ \hline
\end{tabular}
\end{table}

\newpage
\clearpage
\begin{turnpage}
\begin{table}
\caption{The coefficients of selected HCTH/147 basis set fitted functionals \label{tab2}.}
\begin{tabular}{|l|l|l|l|l|l|l|l|l|} \hline
Coefficients         & @3-21G & @6-31+G** & @DZP & @aug-cc-pVDZ & @6-311+G(3d,2p) & @TZ2P & @aug-cc-pVTZ & @cc-pVQZ \\ \hline\hline
$c_1=c_{X\sigma,0}$         &  0.96269 &  1.09617 &  1.10728 &  1.09903 &  1.09434 &  1.09025 &  1.08694 &  1.08782 \\ \hline
$c_2=c_{C\sigma\sigma,0}$   &  5.46039 &  0.86360 &  1.16435 &  1.00558 &  0.35653 &  0.56258 &  0.43359 &  0.47682 \\ \hline
$c_3=c_{C\alpha\beta,0}$    &  1.35755 &  0.61811 &  0.53857 &  0.36506 &  0.48129 &  0.54235 &  0.55240 &  0.46567 \\ \hline
$c_4=c_{X\sigma,1}$         &  1.01633 & -0.61654 & -1.07061 & -0.68459 & -0.68022 & -0.79919 & -0.52152 & -0.67853 \\ \hline
$c_5=c_{C\sigma\sigma,1}$   &  1.64966 & -0.65861 & -2.45161 & -0.71913 &  0.60460 &  0.01714 & -0.07142 &  0.29697 \\ \hline
$c_6=c_{C\alpha\beta,1}$    &  4.77788 &  5.02901 &  7.21706 &  8.78171 &  6.79769 &  7.01464 &  6.31972 &  7.21549 \\ \hline
$c_7=c_{X\sigma,2}$         & -1.90208 &  3.87419 &  5.96561 &  4.42688 &  5.00918 &  5.57212 &  3.77129 &  4.78445 \\ \hline
$c_{8}=c_{C\sigma\sigma,2}$ & -10.334  & -0.3721  &  5.4245  & -1.2823  & -3.8674  & -1.3063  & -1.1795  & -2.1521  \\ \hline
$c_{9}=c_{C\alpha\beta,2}$  & -49.342  & -16.913  & -28.774  & -30.845  & -24.128  & -28.382  & -16.407  & -22.881  \\ \hline
$c_{10}=c_{X\sigma,3}$      &  11.314  & -1.4469  & -7.5266  & -2.7257  & -3.8054  & -5.8676  & -0.3338  & -3.5355  \\ \hline
$c_{11}=c_{C\sigma\sigma,3}$&  3.6314  & -1.0619  & -9.2072  &  2.8592  &  5.1594  &  1.0575  &  0.3399  &  2.8817  \\ \hline
$c_{12}=c_{C\alpha\beta,3}$ &  77.455  &  11.587  &  37.650  &  34.961  &  22.585  &  35.033  &  3.4357  &  19.866  \\ \hline
$c_{13}=c_{X\sigma,4}$      & -3.7768  &  1.3633  &  5.7849  &  6.1824  &  5.4561  &  3.0454  & -2.1418  &  0.5397  \\ \hline
$c_{14}=c_{C\sigma\sigma,4}$& -5.4840  &  2.2222  &  7.0638  & -1.4161  & -1.5290  &  0.8854  &  1.1322  & -0.6523  \\ \hline
$c_{15}=c_{C\alpha\beta,4}$ & -46.989  & -7.5787  & -24.401  & -20.981  & -11.340  & -20.428  &  0.5629  & -9.7235  \\ \hline
\end{tabular}
\end{table}
\end{turnpage}

\newpage
\clearpage
\begin{table}
\caption{The errors of the HCTH/147@ ``basis set functionals'' evaluated over 147 systems with the 6-31+G**
basis set.\label{tab3}}
\begin{tabular}{|l|l|l|l|} \hline
Property                  & RMS energy & Gradient     & Q$_1$   \\ \hline
Functional                & [kcal/mol] & $\sum$[a.u.] & [a.u.]  \\ \hline\hline
HCTH/147@3-21G            & 70.4       & 6.07         & 1460.2  \\ \hline
HCTH/147@6-31G            & 30.0       & 5.11         & 377.57  \\ \hline
HCTH/147@6-31G**          & 9.1        & 2.55         & 41.3    \\ \hline
HCTH/147@cc-pVDZ          & 10.7       & 2.29         & 46.1    \\ \hline
HCTH/147@DZP              & 7.7        & 2.57         & 33.6    \\ \hline
HCTH/147@aug-cc-pVDZ      & 8.4        & 2.58         & 38.9    \\ \hline
HCTH/147@6-311G(2d,p)     & 6.4        & 3.39         & 41.1    \\ \hline
HCTH/147@cc-pVTZ          & 6.9        & 3.58         & 45.3    \\ \hline
HCTH/147@TZ2P             & 6.9        & 3.63         & 46.8    \\ \hline
HCTH/147@6-311+G(2d,2p)   & 6.5        & 3.51         & 43.7    \\ \hline
HCTH/147@6-311+G(3d,2p)   & 6.7        & 3.67         & 46.5    \\ \hline
HCTH/147@6-311+G(3df,3pd) & 6.8        & 3.89         & 50.3    \\ \hline
HCTH/147@aug-cc-pVTZ      & 7.0        & 3.94         & 58.0    \\ \hline
HCTH/147@cc-pVQZ          & 7.5        & 3.90         & 52.6    \\ \hline
\end{tabular}
\end{table}

\newpage
\clearpage
\begin{table}
\caption{The errors of the HCTH/147@ ``basis set functionals'' evaluated over 147 systems with the
aug-cc-pVDZ basis set.\label{tab4}}
\begin{tabular}{|l|l|l|l|} \hline
Property                  & RMS energy & Gradient     & Q$_1$   \\ \hline
Functional                & [kcal/mol] & $\sum$[a.u.] & [a.u.]  \\ \hline\hline
HCTH/147@3-21G            & 70.8       & 4.88         & 1380.8  \\ \hline
HCTH/147@6-31G            & 32.6       & 5.65         & 536.2   \\ \hline
HCTH/147@6-31G**          & 9.3        & 3.40         & 52.3    \\ \hline
HCTH/147@cc-pVDZ          & 9.7        & 3.15         & 49.8    \\ \hline
HCTH/147@DZP              & 8.3        & 3.40         & 48.4    \\ \hline
HCTH/147@6-31+G**         & 7.2        & 3.49         & 45.5    \\ \hline
HCTH/147@6-311G(2d,p)     & 6.3        & 4.16         & 52.1    \\ \hline
HCTH/147@cc-pVTZ          & 6.9        & 4.31         & 56.9    \\ \hline
HCTH/147@TZ2P             & 8.3        & 4.37         & 64.0    \\ \hline
HCTH/147@6-311+G(2d,2p)   & 6.5        & 4.21         & 54.9    \\ \hline
HCTH/147@6-311+G(3d,2p)   & 8.3        & 4.37         & 64.5    \\ \hline
HCTH/147@6-311+G(3df,3pd) & 7.3        & 4.54         & 63.6    \\ \hline
HCTH/147@aug-cc-pVTZ      & 7.3        & 4.61         & 65.1    \\ \hline
HCTH/147@cc-pVQZ          & 7.8        & 4.58         & 65.6    \\ \hline
\end{tabular}
\end{table}

\newpage
\clearpage
\begin{table}
\caption{The errors of the HCTH/147@ ``basis set functionals'' evaluated over 147 systems with the
6-311+G(3d,2p) basis set.\label{tab5}}
\begin{tabular}{|l|l|l|l|} \hline
Property                  & RMS energy & Gradient     & Q$_1$   \\ \hline
Functional                & [kcal/mol] & $\sum$[a.u.] & [a.u.]  \\ \hline\hline
HCTH/147@3-21G            & 73.3       & 4.90         & 1452.7  \\ \hline
HCTH/147@6-31G            & 28.8       & 6.02         & 432.7   \\ \hline
HCTH/147@6-31G**          & 11.0       & 1.65         & 40.17   \\ \hline
HCTH/147@cc-pVDZ          & 13.0       & 1.73         & 53.7    \\ \hline
HCTH/147@DZP              & 7.7        & 1.69         & 24.4    \\ \hline
HCTH/147@6-31+G**         & 7.2        & 1.65         & 21.3    \\ \hline
HCTH/147@aug-cc-pVDZ      & 9.5        & 1.60         & 31.6    \\ \hline
HCTH/147@6-311G(2d,p)     & 6.9        & 1.89         & 20.9    \\ \hline
HCTH/147@cc-pVTZ          & 6.1        & 1.99         & 18.7    \\ \hline
HCTH/147@TZ2P             & 4.8        & 2.03         & 15.5    \\ \hline
HCTH/147@6-311+G(2d,2p)   & 5.6        & 1.96         & 16.9    \\ \hline
HCTH/147@6-311+G(3df,3pd) & 5.0        & 2.19         & 17.0    \\ \hline
HCTH/147@aug-cc-pVTZ      & 5.1        & 2.25         & 18.0    \\ \hline
HCTH/147@cc-pVQZ          & 5.4        & 2.20         & 18.1    \\ \hline
\end{tabular}
\end{table}

\newpage
\clearpage
\begin{table}
\caption{The errors of the HCTH/147@ ``basis set functionals'' evaluated over 147 systems with the
6-311+G(3d,2p) basis set.\label{tab6}}
\begin{tabular}{|l|l|l|l|} \hline
Property                  & RMS energy & Gradient     & Q$_1$   \\ \hline
Functional                & [kcal/mol] & $\sum$[a.u.] & [a.u.]  \\ \hline\hline
HCTH/147@3-21G            & 73.9       & 4.88         & 1415.9  \\ \hline
HCTH/147@6-31G            & 29.6       & 5.29         & 340.9   \\ \hline
HCTH/147@6-31G**          & 11.8       & 1.72         & 45.0    \\ \hline
HCTH/147@cc-pVDZ          & 12.8       & 1.77         & 52.0    \\ \hline
HCTH/147@DZP              & 7.4        & 1.72         & 23.4    \\ \hline
HCTH/147@6-31+G**         & 7.8        & 1.67         & 23.9    \\ \hline
HCTH/147@aug-cc-pVDZ      & 9.8        & 1.63         & 33.0    \\ \hline
HCTH/147@6-311G(2d,p)     & 6.9        & 1.96         & 21.6    \\ \hline
HCTH/147@cc-pVTZ          & 5.8        & 2.02         & 18.7    \\ \hline
HCTH/147@6-311+G(2d,2p)   & 6.3        & 1.98         & 20.2    \\ \hline
HCTH/147@6-311+G(3d,2p)   & 5.6        & 2.09         & 18.6    \\ \hline
HCTH/147@6-311+G(3df,3pd) & 5.9        & 2.23         & 20.5    \\ \hline
HCTH/147@aug-cc-pVTZ      & 5.9        & 2.26         & 21.1    \\ \hline
HCTH/147@cc-pVQZ          & 5.6        & 2.28         & 19.5    \\ \hline
\end{tabular}
\end{table}

\newpage
\clearpage
\begin{table}
\caption{The errors of selected hybrid functionals evaluated for the 147 Set.\label{tab7}}
\begin{tabular}{|l|l|l|l|l|} \hline
Property                 & RMS energy & Gradient     & Q$_1$  & \% HF  \\ \hline
Functional               & [kcal/mol] & $\sum$[a.u.] & [a.u.] & exchange \\ \hline\hline
B97-1/147@6-31+G**       & 7.64       & 2.42         & 34.1   &  0     \\ \hline
B97-1/147@aug-cc-pVDZ    & 8.29       & 3.17         & 44.1   &  0     \\ \hline
B97-1/147@6-311+G(3d,2p) & 5.83       & 2.04         & 18.4   &  0     \\ \hline
B97-1/147@TZ2P           & 6.04       & 2.03         & 19.2   &  0     \\ \hline
B97-1/147@aug-cc-pVTZ    & 6.82       & 2.18         & 24.8   &  0     \\ \hline\hline
B97-1/147@6-31+G**       & 4.29       & 1.79         & 15.0   & 30     \\ \hline
B97-1/147@aug-cc-pVDZ    & 4.29       & 2.68         & 24.5   & 27     \\ \hline
B97-1/147@6-311+G(3d,2p) & 2.94       & 1.86         & 10.3   & 15     \\ \hline
B97-1/147@TZ2P           & 3.47       & 1.83         & 12.0   & 18     \\ \hline
B97-1/147@aug-cc-pVTZ    & 3.04       & 2.07         & 13.2   & 18     \\ \hline
\hline
\end{tabular}
\end{table}


\newpage
\clearpage
\begin{table}
\caption{Errors evaluated with the 407 Set of contemporary functionals, using the TZ2P and cc-pVTZ basis
sets. \label{tab8}}
\begin{tabular}{|l|l|l|l|l|l|l|} \hline
Basis Set          &\multicolumn{3}{c|}{TZ2P}          &\multicolumn{3}{c|}{cc-pVTZ} \\ \hline
Functional         & RMS energy & Gradient     & Q$_1$ & RMS energy & Gradient     & Q$_1$ \\ \hline
HF                 & 155        & 34.81        & 17917 &\multicolumn{3}{c|}{}        \\ \hline
LSDA               & 105        & 15.93        & 7846  &\multicolumn{3}{c|}{}        \\ \hline
PKZB               & 18.0       & 20.45        & 428.6 &\multicolumn{3}{c|}{}        \\ \hline
PBE                & 20.7       & 15.68        & 426.0 & 20.5       & 16.42        & 449.5 \\ \hline
mPW1K              & 16.8       & 19.18        & 399.6 & 17.7       & 18.27        & 483.2 \\ \hline
PW91PW91           & 19.2       & 14.64        & 373.1 & 19.1       & 15.08        & 378.3 \\ \hline
BP86               & 16.9       & 16.16        & 338.2 & 11.8       & 16.51        & 331.4 \\ \hline
mPW91PW91          & 13.8       & 15.66        & 274.0 & 13.9       & 15.41        & 263.4 \\ \hline
BLYP               & 9.8        & 18.50        & 249.7 & 11.0       & 18.91        & 283.9 \\ \hline
BPW91              & 10.3       & 15.51        & 203.1 & 10.9       & 16.80        & 258.4 \\ \hline
PBE0               & 11.9       & 11.19        & 199.8 & 12.2       & 12.05        & 219.8 \\ \hline
OLYP               & 9.6        & 13.52        & 172.5 & 10.0       & 14.32        & 205.5 \\ \hline
B98                & 8.7        & 13.40        & 166.2 & 8.9        & 13.39        & 175.7 \\ \hline
B3LYP              & 9.6        & 11.36        & 165.3 & 10.2       & 11.62        & 177.3 \\ \hline
B97-2              & 7.4        & 11.50        & 161.6 & 8.4        & 11.77        & 175.7 \\ \hline
VSXC               & 9.4        & 11.39        & 158.6 & 9.4        & 11.43        & 167.5 \\ \hline
HCTH/147           & 9.1        & 11.37        & 137.3 & 9.5        & 12.36        & 173.3 \\ \hline
HCTH/407           & 8.0        & 11.28        & 135.3 & 9.3        & 12.46        & 187.6 \\ \hline
HCTH/93            & 8.4        & 11.66        & 134.2 & 9.8        & 12.50        & 178.6 \\ \hline
B97-1              & 7.3        & 10.81        & 130.9 & 8.1        & 11.19        & 143.7 \\ \hline
$\tau$-HCTH        & 7.3        & 10.65        & 114.2 & 8.4        & 11.72        & 150.4 \\ \hline
$\tau$-HCTH hybrid & 6.3        & 10.36        & 107.5 & 7.3        & 11.11        & 133.2 \\ \hline
\end{tabular}
\end{table}


\newpage
\clearpage
\begin{table}
\caption{Number of molecules in the 407 set with an energy error larger than 15 kcal/mol for the functionals tested,
using the TZ2P and cc-pVTZ basis sets. \label{tab9}}
\begin{tabular}{|l|l|l|} \hline\hline
Functional         & TZ2P      & cc-pVTZ  \\ \hline
PBE                & 159       & 140      \\ \hline
BP86               & 130       & 116      \\ \hline
PBE0               & 50        & 48       \\ \hline
BPW91              & 44        & 48       \\ \hline
BLYP               & 36        & 46       \\ \hline
OLYP               & 34        & 46       \\ \hline
B3LYP              & 31        & 37       \\ \hline
HCTH/147           & 30        & 38       \\ \hline
VSXC               & 30        & 33       \\ \hline
HCTH/93            & 24        & 38       \\ \hline
HCTH/407           & 23        & 38       \\ \hline
$\tau$-HCTH        & 22        & 30       \\ \hline
B98                & 21        & 24       \\ \hline
B97-1              & 18        & 23       \\ \hline
$\tau$-HCTH hybrid & 14        & 21       \\ \hline
\end{tabular}
\end{table}


\begin{thebibliography}{99}
\bibitem{PBE} J. P. Perdew, K. Burke, and M. Ernzerhof, \textrm{Phys. Rev. Lett.} {\bf 77}, 3865 (1996).
\bibitem{mPW91} C. Adamo and V. Barone, \textrm{J. Chem. Phys.} {\bf 108}, 664 (1998).
\bibitem{VSXC} T. Van Voorhis, and G. E. Scuseria, J. Chem. Phys {\bf 109}, 400 (1998).
\bibitem{PBE0} C. Adamo and V. Barone, \textrm{Chem. Phys. Lett.} {\bf 298}, 113 (1998).
\bibitem{PKZB} J. P. Perdew, S. Kurth, A. Zupan and P. Blaha, Phys. Rev. Lett {\bf 82} 2544 (1999)
\bibitem{HCTH93} F. A. Hamprecht, A. J. Cohen, D. J. Tozer and N. C. Handy, \textrm{J. Chem. Phys.} {\bf 109}, 6264 (1998).
\bibitem{HCTH120} A. D. Boese, N. Doltsinis, N. C. Handy, and M. Sprik, \textrm{J. Chem. Phys.} {\bf 112} 1670 (2000).
\bibitem{HCTH407} A. D. Boese and N. C. Handy, \textrm{J. Chem. Phys.} {\bf 114} 5497 (2001).
\bibitem{OPTX} N. C. Handy and A. J. Cohen, \textrm{Mol. Phys.} {\bf 99}, 403 (2001).
\bibitem{B972} P. J. Wilson, T. J. Bradley, and D. J. Tozer, \textrm{J. Chem. Phys.} {\bf 115}, 9233 (2001).
\bibitem{tHCTH} A. D. Boese and N.C. Handy, \textrm{J. Chem. Phys.} {\bf 116}, 9559 (2002).
\bibitem{mPW1K} B. J. Lynch, P. L. Fast, M. Harris, and D. G. Truhlar \textrm{J. Phys. Chem. A} {\bf 104}, 21 (2000).
\bibitem{B97} A. D. Becke, \textrm{J. Chem. Phys.} {\bf 107}, 8554 (1997).
\bibitem{B98} A. D. Becke, \textrm{J. Chem. Phys.} {\bf 109}, 2092 (1998).
\bibitem{B88X} A. D. Becke, \textrm{Phys. Rev. A} {\bf 38} 3098 (1988).
\bibitem{B3P91} A. D. Becke, \textrm{J. Chem. Phys.} {\bf 98} 5648 (1993).
\bibitem{VWN} S. J. Vosko, L. Wilk and M. Nusair, \textrm{Can. J. Phys.} {\bf 58} 1200 (1980)
\bibitem{P86} J. P. Perdew, \textrm{Phys. Rev. B.} {\bf 33} 8822 (1986).
\bibitem{LYP} C. Lee, W. Yang, R. G. Parr, \textrm{Phys. Rev. B} {\bf 37} 785 (1988).
\bibitem{P91X} J. P. Perdew and Y. Wang, \textrm{Phys. Rev. B.} {\bf 45} 13244 (1992)
\bibitem{P91c} J. P. Perdew, J. A. Chevary, S. H. Vosko, K. A. Jackson, M. R. Pederson, D. J. Singh, C. Fiolhais, \textrm{Phys. Rev. B.} {\bf 46} 6671 (1992).
\bibitem{basquad} F. De Proft, F. Tielens, and P. Geerlings, \textrm{J. Mol. Struct. (Theochem)} {\bf 506}, 1 (2000).
\bibitem{basquad2}  F. De Proft, J. M. L. Martin, and P. Geerlings, \textrm{Chem. Phys. Lett} {\bf 256}, 400 (1996).
\bibitem{basdip} A. C. Scheiner, J. Baker, and J. W. Andzelm, \textrm{J. Comp. Chem.} {\bf 18}, 775 (1997).
\bibitem{basat1} J. M. Martell, J. D. Goddard, and L. A. Eriksson, \textrm{J. Phys. Chem. A} {\bf 101}, 1927 (1997).
\bibitem{basblyp} P. M. W. Gill, B. G. Johnson, J. A. Pople, and M. J. Frisch, \textrm{Chem. Phys. Lett.} {\bf 197}, 499 (1992).
\bibitem{basfreq} J. M. L. Martin, J. El-Yazal, and J. P. Fran\c{c}ois, \textrm{Mol. Phys.} {\bf 86}, 1437 (1995).
\bibitem{basrev} J. M. L. Martin, in ``Density Functional Theory : a bridge between Chemistry and Physics'' (P. Geerlings, F. De Proft, and W. Langenaeker, eds.), VUB Press, Brussels, 2000, pp. 111.
\bibitem{Jensen1} F. Jensen, \textrm{J. Chem. Phys.}, {\bf 115}, 9113 (2001).
\bibitem{Jensen2} F. Jensen, \textrm{J. Chem. Phys.}, {\bf 116}, 7372 (2002).
\bibitem{GSAW} N. Godbout, D. R. Salahub, J. Anzelm, and E. Wimmer, \textrm{Can. J. Chem.}, {\bf 70}, 560 (1992).
\bibitem{DFO} D. Porezag and M. R. Pederson, \textrm{Phys. Rev. A}, {\bf 60}, 2840 (1999).
\bibitem{G2-1}  L. A. Curtiss, K. Raghavachari, G. W. Trucks, and J. A. Pople, \textrm{J. Chem. Phys.} {\bf 94}, 7221 (1991).
\bibitem{G3} L. A. Curtiss, K. Raghavachari, P. C. Redfern, and J. A. Pople, \textrm{J. Chem. Phys.} {\bf 112}, 7374 (2000).
\bibitem{www} http://www-theor.ch.cam.ac.uk/people/nch/fit/; see also supplementary
material to \cite{HCTH407}, EPAPS Document No. E-JCPSA6-301111; please note that the geometry of benzene in the latter document contained an error.
\bibitem{HCTH407+} A. D. Boese, J. M. L. Martin, A. Chandra, D. Marx, and N. C. Handy, unpublished results
\bibitem{PW91} J. P. Perdew and Y. Wang, \textrm{Phys. Rev. B.} {\bf 45} 13244 (1992)
\bibitem{ZMP} Q. Zhao, R. C. Morrison, and R. G. Parr, \textrm{Phys. Rev. A} {\bf 50}, 2138 (1994).
\bibitem{TZ2P} T. H. Dunning, \textrm{J. Chem. Phys.} {\bf 55}, 716 (1971).
\bibitem{PVXZfirst} R. A. Kendall, T. H. Dunning, and R. J. Harrison, \textrm{J. Chem. Phys.} {\bf 96}, 6796 (1992).
\bibitem{PVXZsecond} D. E. Woon and T. H. Dunning, \textrm{J. Chem. Phys.} {\bf 99}, 3730 (1993).
\bibitem{Cadpac}
The Cambridge Analytic Derivatives Package (Cadpac), Issue 6.5, Cambridge, 1998
Developed by R. D. Amos with contributions from I. L. Alberts,
J. S. Andrews, S. M. Colwell, N. C. Handy, D. Jayatikala, P.J. Knowles,
R. Kobayashi,K. E. Laidig, G. Laming, A. M. Lee, P. E. Maslen,
C. W. Murray, P. Palmieri, J. E. Rice, E. D. Simandiras, A. J. Stone,
M.-D. Su, and D. J. Tozer.
\bibitem{321G1} J. S. Binkley, J. A. Pople, and W. J. Hehre, \textrm{J. Am. Chem. Soc.} {\bf 102}, 939 (1980).
\bibitem{321G2} M. S. Gordon, J. S. Binkley, J. A. Pople, W. J. Pietro, and W. J. Hehre, \textrm{J. Am. Chem. Soc.} {\bf 104}, 5039 (1982).
\bibitem{631G1} W. J. Ditchfield and J. A. Pople, \textrm{J. Chem. Phys.} {\bf 56}, 2257 (1971).
\bibitem{631G2} P. C. Hariharan and J. A. Pople, \textrm{Theor. Chim. Acta} {\bf 28}, 213 (1973).
\bibitem{6311G1} M. J. Frisch, J. A. Pople, and J. S. Binkley, \textrm{J. Chem. Phys.} {\bf 80}, 3265 (1984).
\bibitem{6311G2} R. Krishnan, J. S. Binkley, R. Seeger, and J. A. Pople, \textrm{Theor. Chem. Acc.} {\bf 72}, 650 (1980).
\bibitem{BD} R. A. Chiles and C. E. Dykstra, \textrm{J. Chem. Phys.} {\bf 74}, 4544 (1981).
\bibitem{CH} A. J. Cohen and N. C. Handy, \textrm{Chem. Phys. Lett.} {\bf 316}, 160 (2000).
\bibitem{CHcomment} R. Ahlrichs, F. Furche, and S. Grimme, \textrm{Chem. Phys. Lett.} {\bf 325}, 317 (2000).
\bibitem{Menconi} G. Menconi, and D. J. Tozer, \textrm{Chem. Phys. Lett.} {\bf 360}, 38 (2002).
\bibitem{Menconif} G. Menconi, P. J. Wilson, and D. J. Tozer, \textrm{J. Chem. Phys.} {\bf 114}, 3958 (2001).
\bibitem{CWNZ} K. A. Peterson and T. H. Dunning, Jr., \textrm{J. Chem. Phys.} {\bf 117}, 10548 (2002).
\bibitem{MartinCV} M. A. Iron, M. Oren, and J. M. L. Martin, \textrm{Mol. Phys.}, accepted for publication.
\bibitem{WilsonD} T. H. Dunning, Jr., K. A. Peterson, and A. Wilson, \textrm{J. Chem. Phys.} {\bf 114}, 9244 (2001).
\bibitem{SO2martin} J. M. L. Martin, \textrm{J. Chem. Phys.} {\bf 108}, 2791 (1998).
\bibitem{anharm} A. D. Boese and J. M. L. Martin, unpublished results.
\bibitem{TBH} C. Tuma, A. D. Boese, and N. C. Handy, \textrm{Phys. Chem. Chem. Phys.} {\bf 1}, 3939 (1999); note that the HCTH/120 functional is denoted HCTH-38 in this paper.
\bibitem{BLYPCP} M. Sprik, J. Hutter, and M. Parrinello, \textrm{J. Chem. Phys.} {\bf 105}, 1142 (1996).
\bibitem{OLYP} N. C. Handy, and A. J. Cohen, \textrm{Mol. Phys.} {\bf 99}, 403 (2001).
\bibitem{hoe} W. M. Hoe, A. J. Cohen, and N. C. Handy \textrm{Chem. Phys. Lett.} {\bf 341}, 319 (2001).
\bibitem{baker} J. Baker and P. Pulay, \textrm{J. Chem. Phys.} {\bf 117}, 1441 (2002).
\bibitem{BHunpub} A. D. Boese and N. C. Handy, unpublished results.
\bibitem{mPW1Keval} M. Harris and D. G. Truhlar \textrm{J. Phys. Chem. A} {\bf 105}, 2936 (2001).
\bibitem{W2} J. M. L. Martin and G. De Oliveira, \textrm{J. Chem. Phys.} {\bf 111}, 1843 (1999).

\end{thebibliography}
\end{document}